\documentclass{aa} % for a referee version
\usepackage{graphics}
\usepackage{psfig}
\def\la{\raise.5ex\hbox{$<$}\kern-.8em\lower 1mm\hbox{$\sim$}}
\def\ma{\raise.5ex\hbox{$>$}\kern-.8em\lower 1mm\hbox{$\sim$}}

\def\msol{M$_{\odot}$ }
\def\kms{$\rm km\, s^{-1}$}
\def\cm3{$\rm cm^{-3}$}
\def\Vs{$V_{\rm s}$}
\def\n0{$n_{\rm 0}$}
\def\B0{$B_{\rm 0}$}

\def\erg{$\rm erg \, cm^{-2} \, s^{-1}$}

\def\Hb{H$\beta$}

\def\kms{$\rm km \, s^{-1}$}
\def\cm3{$\rm cm^{-3}$}

\def\Vs{$\rm V_{s}$}
\def\n0{$\rm n_{0}$}
\def\B0{$\rm B_{0}$}
\def\Fh{$\rm F_{H}$}

\def\Hb{H$\beta$}

\def\erg{$\rm erg \, cm^{-2} \, s^{-1}$}

\begin{document}

   \thesaurus{} 

\title{The signature of high velocity gas in the spectra of NGC 4151}

\author{
M. Contini\inst{1,2},
S. M. Viegas\inst{2}, and M.A. Prieto\inst{3}
}

\offprints{M.\ Contini, contini@ccsg.tau.ac.il}

   \institute{
School of Physics \& Astronomy, Tel Aviv University,
69978 Tel Aviv, Israel
\and
Instituto Astron\^{o}mico e Geof\'{i}sico, USP, Av. Miguel Stefano,
4200,04301-904 S\~{a}o Paulo, Brazil
\and
European Southern Observatory, D-85748 Garching, Germany}

   \date{Received ??; accepted ??}

\authorrunning{M.\ Contini, S.M.\ Viegas, and M.A.\ Prieto}
\titlerunning{High velocity gas in NGC 4151}
   \maketitle

\begin{abstract}

The multiwavelength emission spectrum and associated velocity field
of the  Seyfert prototype NGC 4151 is modeled.
NGC 4151 has been since the thirties subject of extensive modeling,
nuclear photoionization being the basic approach considered by all
authors.
HST data has impressively revealed the
existence of a large range of velocities (100 - 1500 \kms)
dominating the emitting clouds in the
extended emission line region of the galaxy. Following this
observational result, a revision of the  photoionization modeling
approach
applied to  NGC 4151 is presented.
It is concluded that a mixture of radiation dominated clouds and shock
dominated clouds are required to explain the multiwavelength
line and continuum  spectra of the galaxy.
The relative contribution of shock excitation versus photoionization
is consistently  modeled along the nebulae
taking into account the spatial variation of both flux and  velocity of
main
optical and UV lines.
The multiwavelength continuum spectrum of NGC 4151 is then nicely
accounted for
by a combination of nuclear emission at high energy, gas Bremsstrahlung,
 and dust emission. The last two phenomena are
 directly linked  to  the composite shock and photoionization
excitation of the  gas.
The radio SED is found however dominated by
synchrotron emission  created by Fermi mechanism at the shock front.
In addition, a 3x10$^3$ K black body component
accounts for the host  galaxy contribution.
As a result of the modeling,
silicon is found  depleted  by a
high factor and included in dust
grains, while N/C abundance ratio is found compatible with cosmic
values.


\end{abstract}

%%%%%%%%%%%%%%%%%%%%%%%%%%%%%%%%%%%%%%%%%%%%%%%%%%%%%%%%%%%%%%%%%%%%%%%%%%%%%%%%

\section{Introduction}

Detailed imaging of the narrow-line region (NLR) of Seyfert galaxies
with the {\it Hubble Space Telescope} (HST) has revealed its complex
morphology
and  velocity field. Recently, observations of NGC 4151
indicated the presence of emitting clouds with velocities ranging
from +846 to -1716  \kms (Winge et al. 1997, Kaiser et al. 1999).

NGC 4151 is a nearby barred galaxy,  usually classified as a Seyfert
type 1, although it has already been considered as a Seyfert 1.5
(Osterbrock \& Koski 1976) and has also shown the
characteristics of a Seyfert 2 galaxy (Penston \& P\'erez 1984). NGC
4151
is one of the most observed active galactic nuclei (AGN), from radio to
X-rays. The recent HST longslit emission-line data (Nelson et al.
2000), coupled to ISO data  (Alexander et al.  1999),
offer an excelent opportunity to improve our understanding of the
NLR physical conditions with  the added  effect of the velocity
field.

The kinematics derived from the long slit observations (Nelson et al.
2000, Crenshaw et al. 2000) show evidence of three components: a low
velocity system,
consistent with normal
disk rotation, a high velocity system in radial outflow at a few hundred
\kms and an additional high velocity system with velocities up to
1400 \kms, as previously found from STIS slitless spectroscopy
(Hutchings
et al. 1998, 1999, Kaiser et al. 1999). The authors see the signature of
a radial outflow, with no interaction with the radio jet. However,
high velocity components, shifted up to about 1500 \kms ~from the
systemic velocity, are also  seen
by Winge et al (1997) associated with individual clouds
located preferentially along the edges of the radio knots. Such
association suggests a cloud-jet interaction, which definitively may
 influence the morphology
and the  physical conditions of the NLR.

The high spatial resolution permits to see  that the emission
line ratios vary substantially on scales
of a few tenth of an arcsecond, indicating that the density and
ionization
state of the emitting gas are strongly influenced by the local
conditions, hence
suggesting that shock fronts may be at work.

The presence of high velocity clouds in the NLR of active galaxies
has been predicted from the  modeling
of the emission-line and continuum spectra of the Seyfert 2 objects
NGC 5252  and Circinus  (Contini, Prieto \& Viegas 1998a, 1998b), as
well as  suggested from  the more general
analysis of the optical-ultraviolet
continuum of a relatively large sample of Seyfert 2 galaxies (Contini
\& Viegas 2000). High velocity clouds are  also
 invoked in  an alternative model
  based on a multiwavelength analysis of
the  emission-line and continuum spectra of the source
(Contini \& Viegas 1999), which explains the soft X-ray excess
in NGC 4051.

In all of these models the physical
conditions of the high velocity clouds are usually shock-dominated.
The shocks contribute to the high ionization emission-lines and to the
continuum emission, mainly by providing efficient heating of dust,
contributing to reproduce the observed middle-infrared emission, and
to the observed soft X-ray spectrum which originates in the high
temperature post-shock zone.

Yet,  the interpretation of the NGC 4151  emission-line and continuum
spectra
is so far residing exclusively on  photoionization models.
All the authors agree that, given the simplicity of the models,
the observed line ratios can be reproduced.

In NGC 4151 the high velocity clouds are directly observed.
Velocities of $\sim$ 1500 \kms ~could be theoretically  explained
extrapolating to the NLR the arguments about high velocity clouds
raised by Weymann et al. (1997).
 Here, however, in order to obtain a large range of velocities
collisions
are invoked.

Although
photoionization models reproduce within better than a factor of 2 the
most
important line ratios  (Schulz 1988, 1990;
Kraemer et al. 2000; Alexander et al. 1999),
some questions remain still open, as for example
the mere fact of the influence of the velocity field in the cloud
spectrum.
Thus, in this paper we
will look for the signature   of  the high velocity gas  in the
emission-line spectrum, as well as   in the
radio to X-rays continuum spectral energy distribution (SED).

We present a new type of single-cloud modeling based on the spatial
distribution of the observational data in the optical range;
then, a multi-cloud model is proposed to explain the emission lines in
the IR.

As was done previously in  modeling other objects (NGC 5252, Circinus),
 the multi-cloud model issued from the analysis of the emission lines
is then constrained by fitting
 the continuum SED in a large frequency range.
This is possible only with composite models (shock + photoionization),
because
the gas heated only  by the  radiation from the active center cannot
reach temperatures high enough to fit the Bremsstrahlung emission
and dust reradiation in  the large range of the observed frequencies.

A review focused in the successes and problems of photoionization
models  is
presented in  Sect. 2. The composite models, coupling photoionization
and shocks, are described in  Sect. 3. The optical emission-line
spectrum is
discussed in Sect. 4, while the infrared lines  are presented
in Sect. 5, and the results for the ultraviolet lines
appear in Sect. 6. The observed and calculated SED are compared
in Sect. 7, and the conclusions appear in Sect. 8.

\section{A brief review of  photoionization models}

Most recently, photoionization models for the NLR of NGC 4151 have
been proposed by Alexander et al. (1999) and Kraemer et al. (2000).
In both cases, the observed emission-line
ratios are  reproduced within a factor of 2.
Both set of authors use, however, different modeling approach:
the  matter distribution adopted
is different. In   Alexander's et al. (1999) models  (hereinfter
called type A), the best fit is found with a
 single cloud component with a
filling factor less than unity; in  Kraemer's et al. (2000) models
(type B),  a multi- cloud is used, namely a less dense
matter-bound component is added to a dense
radiation-bounded component. We note that Alexander et al.  discarded
this latter solution arguing that a matter-bounded component implies
three free additional parameters, whereas the introduction of a
 filling factor --their case-- implies only one.
Nevertheless, let us recall
that a matter-bounded component favors the high ionization lines,
while  (radiation-bounded) models with filling-factor less than unity
mimic a lower average density, leading to a larger ionized zone and
favoring the low-ionization lines. Thus,  models assuming a filling
factor less than unity are not  a good
physical representation of a clumpy zone; yet, the  spatial
distribution of [SII] ratios across NGC 4151 extended emission
line region reveals instead a
rather clumpy region with regions of higher and lower density at
different points along the HST slit (Nelson et al. 2000).
Regarding both type A and type B models, the authors
agree that "the fit of the line
ratios is good {\it taking into account the simplicity of the models}".

Simplicity has always been a strong argument in favor of
photoionization models applied to nebular regions in AGNs, in addition
to the
undisputable presence of
a strong central radiation source.  However, in Science
it is usually from the
attempt to explain the "imperfections" of a model, (those data  not
explained by it), that a more realistic scenario can be drawn.
With that in mind we list below the discrepancies
between the above proposed  photoionization models and
the observational data of NGC 4151.

Due to the various difficulties to have a self-consistent data
set (data from different epoch, different resolutions and apertures),
Alexander et al. (1999) use several criteria to select the
set of emission-lines to be reproduced by type A models.
One of the criteria excludes lines emitted by ions that
{\it "can be easily ionized by other processes"}. The set of
 emission-lines excluded
includes some usually used in the diagnostic diagrams of nebular gas:
He II 4686, [O III] 4363, [O I] 6300, [N II] 6548+6584 and the
density indicator [SII] 6716, 6731 doublet,
in addition to the high-ionization Fe lines: [Fe X] 6734 and
[Fe XI] 7892. The final set of lines used for modeling, that includes
UV, optical, and IR lines, is reproduced within a factor of 2 by their
best fit models.
Discrepancies larger than two are derived for [Fe VII] 5721, [Fe VII]
6086,
[Ne II] 12.8 and [Ne III] 36.0 which are underpredicted, and for
 [S III] 9069, [Ne V] 14.3, [S III] 18.7,
[Ne V] 24.3, and [S III] 33.5 that are overpredicted.

 Regarding the SED implied by models,
the presence of a {\it big blue bump}
in the ionizing radiation spectrum, peaking at 50 eV, is excluded
by their best-fit model because its effect would be {\it to overproduce
the low-ionization lines and underproduce the high-ionization
lines}. However, this is  the net effect produced by imposing a
filling factor  less than unity, as adopted in their best-fit model.

Kraemer et al. models (type B )  use   HST/STIS low-dispersion long-slit
data at  position angle P.A. = 221$^o$. Thus, their data
set includes UV and optical lines, but no infrared lines that are
the basis of type A models by Alexander et al. (1999).
Their model results are compared with observed emission-line ratios
 derived at  different positions along the slit. The
large majority of the lines  are reproduced within a factor of
2. Divergencies larger than 2 are often found for the high ionization
lines:
 [Fe VII] lines, and the UV lines
N V 1240, [Ne IV] 2423,  and NIV] 1486, although in the latter case
the signal-to-noise is very low. At some locations of  the emitting
nebulae, the calculated C III] 1909, CII] 2326,  [O III] 4363,
and the [S III] lines also largely diverge from the data.
Notice, however, that CIII] and CII] are blended and that,
in the case of the  [S III] lines the discrepancy may be
due to  an  instrumental effect as pointed out by Kraemer et al. (2000).

Several interpretations for the above discrepancies are discussed by
 Kraemer et al. (2000). Our assumption is that those discrepancies may
be
 revealing the presence of an additional ionizing mechanism, which is
 not the dominant process, but that shows its signature through
 particular observational features. This point of view is adopted
 here assuming that the additional mechanism is due to the presence
of shocks.

\section{Single-cloud models}

Faint high velocity emission regions intermingled with brighter
emission clouds are shown in NGC 4151 imaging (Hutchings et al.
1999). We consider this observational fact as  an  indication
that the extended NLR  of NGC 4151
is a mixture of  low velocity radiation-dominated
clouds and high velocity shock-dominated clouds, all
contributing to the emission-line spectrum. Accordingly,
composite models  accounting
for the coupled effect of the central ionizing radiation
and  shock excitation  due to cloud  motions are assumed.
Numerical simulations for single clouds are obtained with
the SUMA code (see, for instance, Viegas \& Contini 1994). Notice that
the simulations apply whether the shocks originate from an interaction
of the emitting clouds with the radio jet or from a radial
outflow of the clouds.

The input parameters are the shock velocity, \Vs, the preshock density,
\n0, the preshock magnetic field, \B0, the ionizing radiation spectrum,
the chemical abundances,
the dust-to-gas ratio by number, d/g,
and the geometrical thickness of the clouds, D.  A power-law,
characterized by the power index $\alpha$ and the
flux, \Fh, at the Lyman limit, reaching the cloud (in units of
cm$^{-2}$ s$^{-1}$ eV$^{-1}$)
is generally adopted.
For all the models,  \B0 = $10^{-4}$ gauss, $\alpha_{UV}$ = 1.5,
and $\alpha_X$=0.4,
and cosmic abundances (Allen 1973) are adopted.
The basic models are calculated with d/g = $10^{-15}$, however,
this value is changed a posteriori  to  better fit the continuum SED.

Shock dominated models (SD) are calculated
 assuming that the effects of the shock prevail on radiation (\Fh=0.).
Radiation dominated models (RD),
however, are  composite, i.e. they account both  for photoionization
and shocks up to \Vs=500 \kms, but photoionization
dominates the physical conditions of the emitting gas.

The grid  of models which  are actually used for modeling is
presented  in an accompanying paper by Contini \& Viegas (2001,
hereafter referred to as CV01).
In the following, calculated emission-line ratios from a selected
number of  models in the grid
are  compared to the HST log-slit optical
data at P.A.= 221$^o$ (Nelson et al. 2000 and Kraemer et al.
2000) and to the ISO
integrated aperture SWS data  by Sturm et al. (1999).
Models are selected  on the basis of the physical conditions of the
emitting gas dictated by the observations, e.g. the
FWHM of the line profiles
for \Vs.

In models which account for shock effects, the density downstream
is determined by compression, which depends on the shock velocity,
 and changes considerably with
distance from the shock front (CV01, Figs. 5a, 6a, and 7a).
Therefore, the preshock density and the shock velocity chosen
define a distribution of the density across the cloud, which
must be adequate to provide a good fit to the density sensitive
lines, e.g., [OII] 3727, [NI] 5200, [NII]  6548 ([SII] 6716+
is not very significant because S can be locked in dust grains).

Moreover, in the NLR the density of the clouds
follows the gradient of the cloud velocity, decreasing
with the distance to the center. For each cloud,
the observed [S II] 6717/6730  line ratio (which does not depend on S/H)
is used as a first test for the choice of \n0 and V$_s$.
Then, the intensity of the ionizing radiation, the physical conditions
calculated by the model, and the geometrical thickness of the cloud
are deduced  from the  line spectrum, as a whole.
The  [O III]/\Hb ~line ratio is an indicator for \Fh.
On the other hand,
wide clouds are optically thick, leading to stronger
low ionization level lines, while narrow clouds
are matter-bounded, with fainter low-ionization lines.
Thus the choice of D is then constrained  by the best fit of
a large number of  line ratios, particularly,
the ratio of the low ionization lines to \Hb.

\newpage
\section{The optical emission-lines}

\centerline{\it Modeling the spatial distribution}

Because of the availability of spatially resolved emission-line spectra
in different regions of NGC 4151, a new modeling approach is followed in
this work.
The most significant  optical emission lines are modeled
accounting for their spatial distribution across the NLR.

In each position we
check  the consistency of the prevailing  models explaining several
line ratios,  including those that are poorly reproduced by the
photoionization models. 

The observed and calculated emission-line intensities relative
to \Hb ~are presented in a series of figures (Figs. 1-4), where
the emission-line ratios
are shown as a function of the projected nuclear distance, including
both the SW data (on the left) and the NE data (on the right) at P.A.=
221$^o$.

In order to determine the intensity of the power-law
radiation flux in the NLR edge closer to the  nucleus
a preliminary estimate is made by comparing model results with
the observed line ratios as  presented in Fig. 1.
Filled squares refer  to Nelson et al. 2000 (not reddening corrected)
 and  filled triangles to
Kraemer et al. 2000 (reddening corrected) data, respectively. 

 The scales for log \Fh
~are indicated upon the  upper horizontal axis.
Each curve  represents RD models calculated with different \Fh ~but
corresponding to
the same  \Vs  - \n0 :
\Vs=100 \kms ~and \n0=100 \cm3 (dotted lines),
\Vs=200 \kms ~and \n0=200 \cm3 (dash-dotted lines),
\Vs=300 \kms ~and \n0=300\cm3 (short-dash lines), \Vs=500 and \n0=300
\cm3, (long-dash lines). Thin lines refer to a narrow cloud
(D=10$^{17-18}$ cm)
and thick lines to wider clouds (D = 10$^{19}$ cm). 

A sequence of three different cases (log
\Fh=13,
12, and 11.3) is shown in the top, middle, and bottom diagrams,
respectively.
 We have chosen the [OIII]/\Hb ~and [OII]/\Hb ~line ratios, because
 are the most  significant.
 The  fit of both   [OIII]/\Hb ~and [OII]/\Hb  ~data is acceptable only
in the bottom
diagrams (notice that in these diagrams the flux is not fully symmetric,
indicating that
the SW and NE regions are slightly different). So, for consistency we
chose the bottom
case to model also the other line ratios.

\begin{figure*}
\psfig{figure=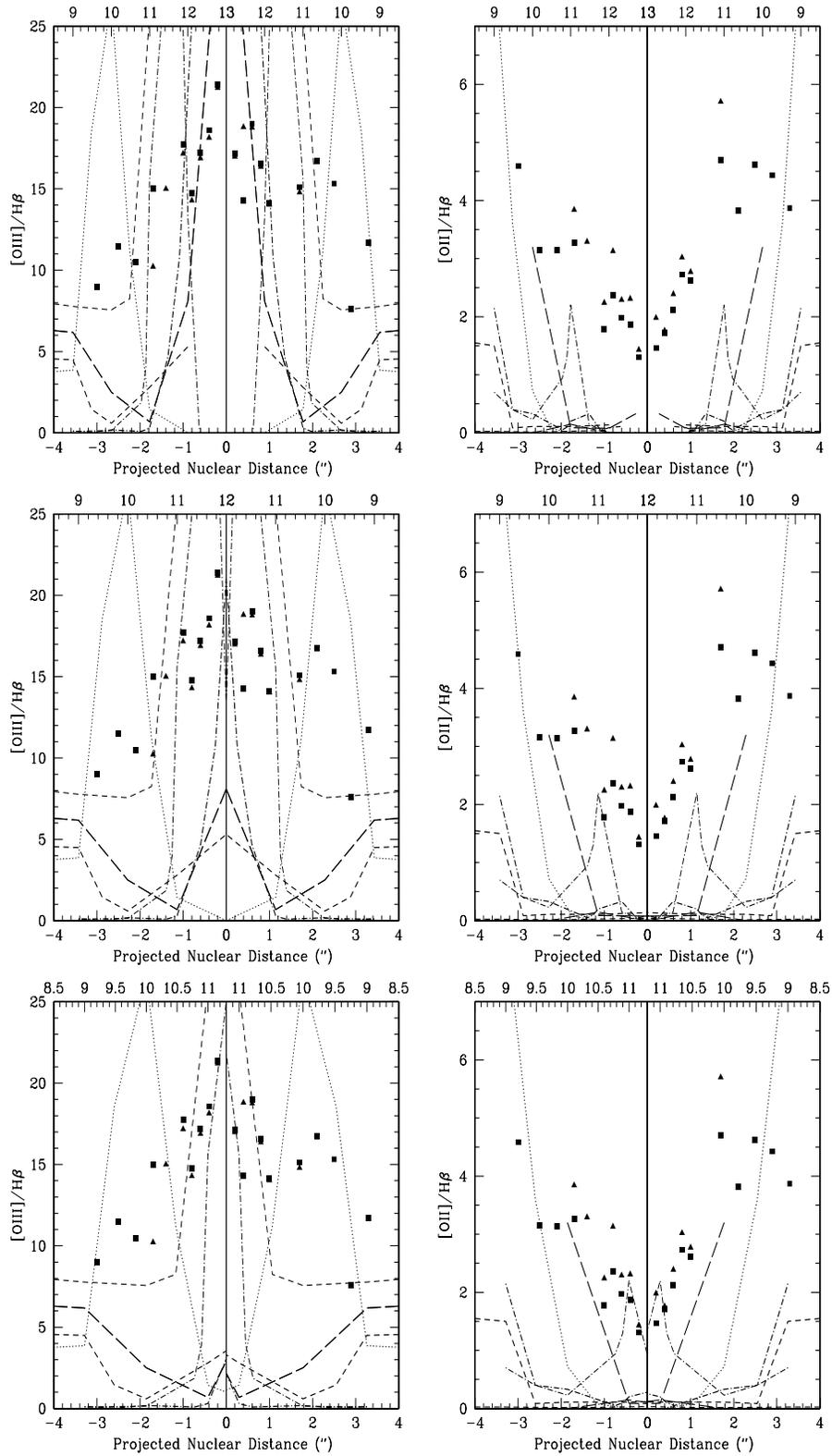,bbllx= 80pt,bblly=
30pt,bburx=490pt,bbury=650pt,width=15.5cm,clip=}
  \caption{The [O III] 5007+4959/\Hb ~(left diagrams) and the [O II] 3727/\Hb
~(right diagrams) emission-line ratios
are shown as a function of the projected nuclear distance.
}
\label {fig:fus1}
\end{figure*}

Since we  are investigating the intensity of ionizing radiation
from the active center (AC)
and the velocity distribution in the central region of NGC 4151, both
\Fh
~and \Vs ~are  shown in  Figs. 2-4  diagrams.
 In this way, the relative importance of the velocity field and the AC
radiation field can be recognized.
 The scales for log \Fh ~and \Vs (in \kms)
~are indicated upon the horizontal upper axis of the
 top and middle and bottom diagrams, respectively.

In the top diagrams  of Figs. 2-4  (panels a and d) each line corresponds to  RD model
results as in Fig. 1.
Panels b and e show
the
solid lines corresponding to SD models (CV01, Tables 1-10)
for which the maximum value of the velocity distribution is
 \Vs=700 \kms ~in the central region, while diagrams c and f correspond
to the
results with a maximum velocity of \Vs=1400 \kms.

Because shock dominated models for
narrow and wide clouds give very similar results,
these models are represented by one line (model results overlap),
corresponding to one serie of results.
We draw attention  to the  thin and thick lines  in the two bottom
diagrams
which refer to reduced and full intensity line ratios,
respectively, and not to models calculated by small and large D, as
shown
in the top diagrams.

\begin{figure*}
\psfig{figure=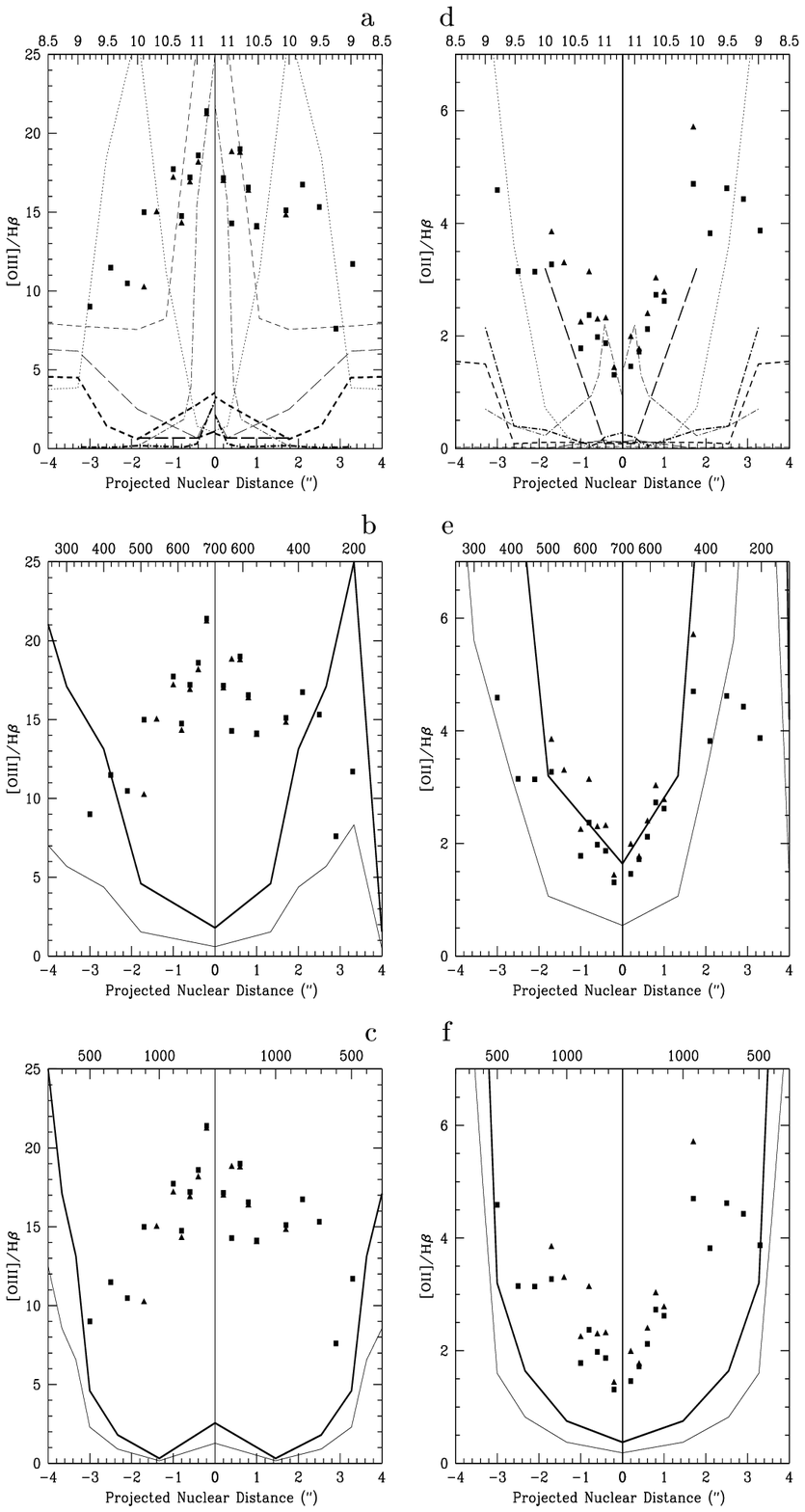,bbllx= 80pt,bblly=
30pt,bburx=490pt,bbury=650pt,width=15.5cm,clip=}
  \caption{The [O III] 5007+4959/\Hb ~(left) and the [O II] 3727/\Hb
~(right)
emission-line ratios
are shown as a function of the projected nuclear distance. Same notation
as in Figure 1.
}
\label {fig:fus1}
\end{figure*}

\begin{figure*}
\psfig{figure=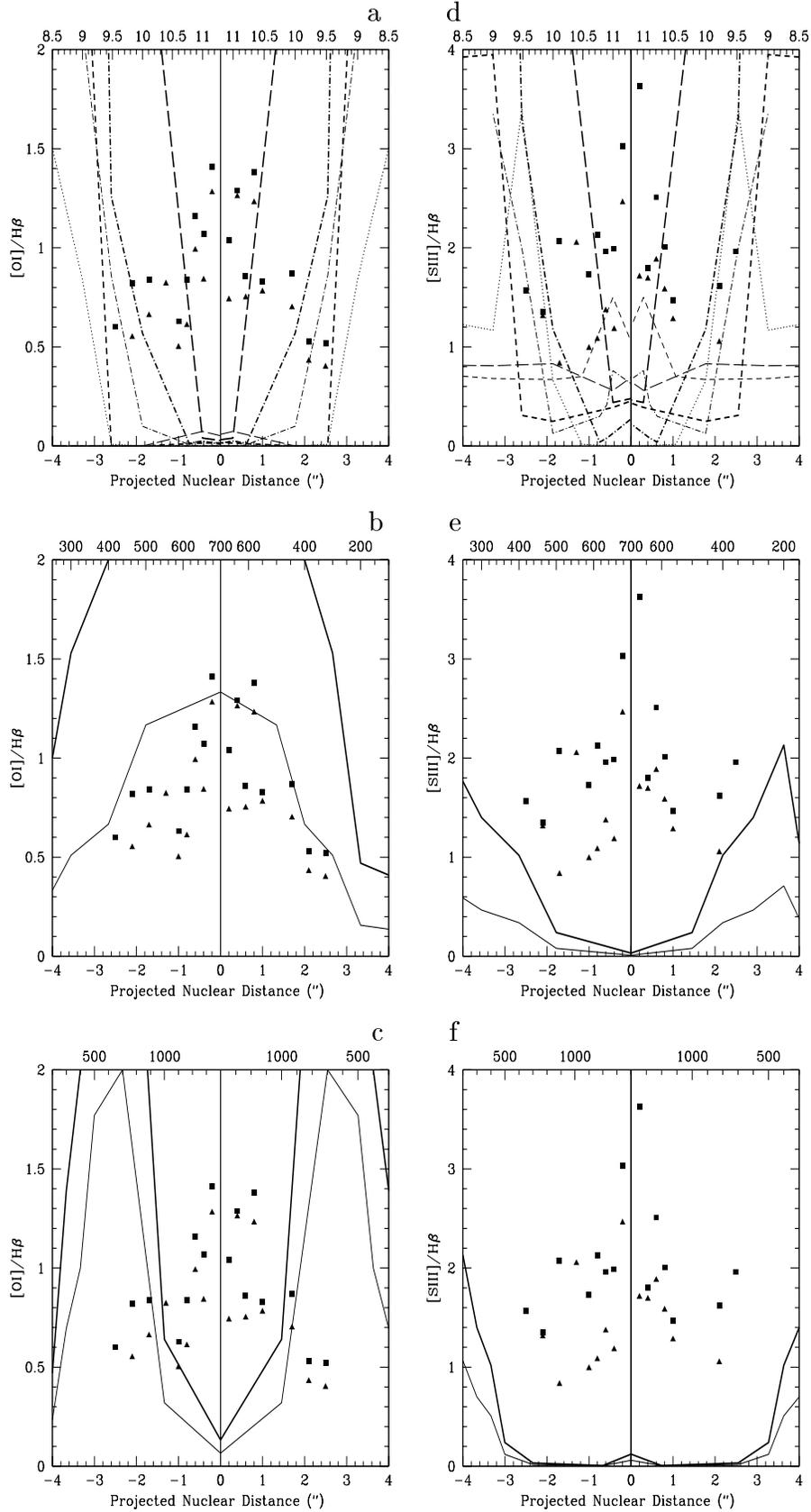,bbllx= 80pt,bblly=
30pt,bburx=490pt,bbury=650pt,width=15.5cm,clip=}
  \caption{The [O I] 6300+6360/\Hb ~(left) and
the [S III] 9069/\Hb ~(right) emission-line ratios as a function of the
projected nuclear distance. Same notation as in Figure 1. 
}
\label {fig:fus2}
\end{figure*}

\begin{figure*}
\psfig{figure=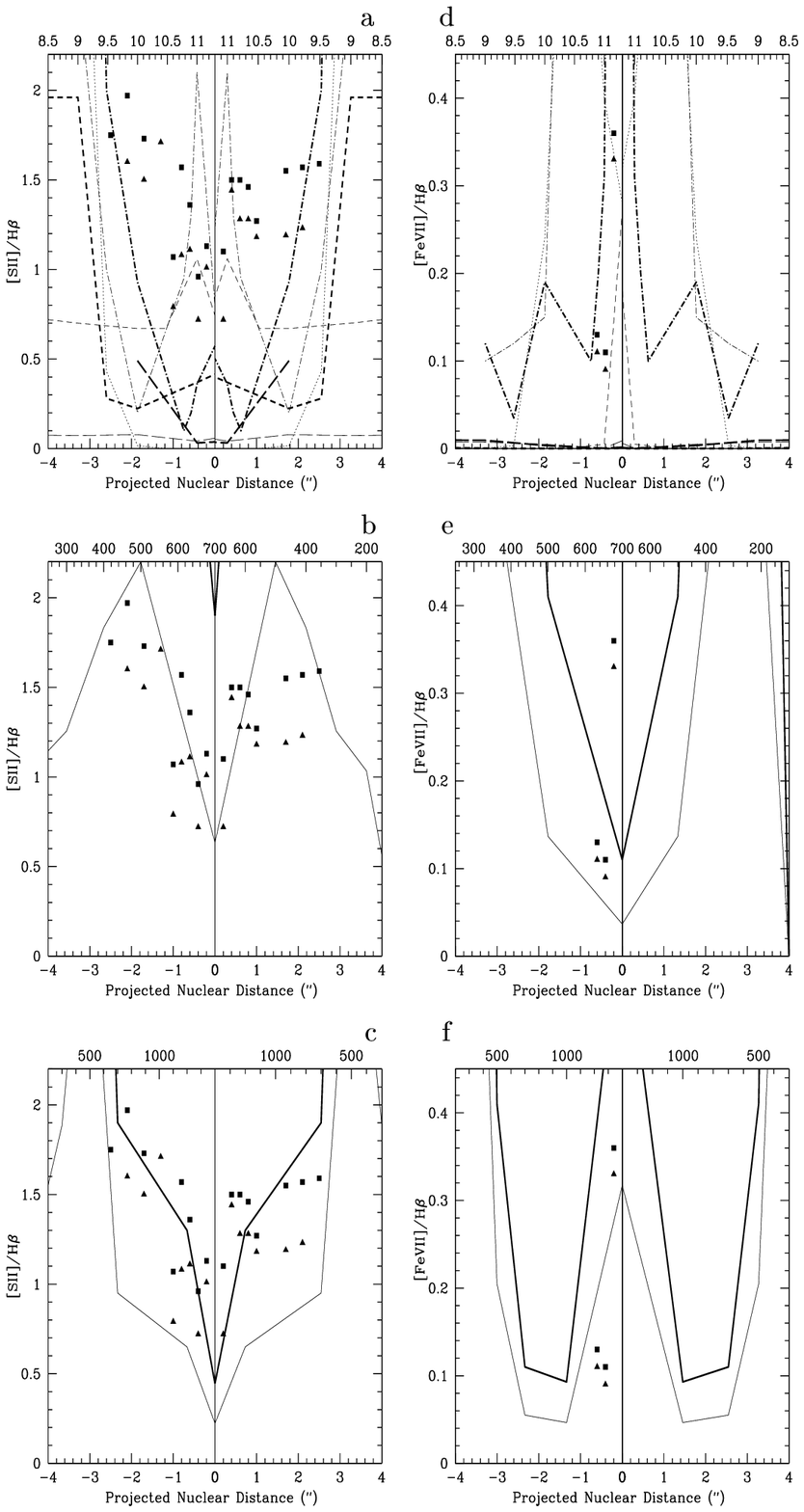,bbllx= 80pt,bblly=
30pt,bburx=490pt,bbury=650pt,width=15.5cm,clip=}
  \caption{The [S II] 6713+6731/\Hb (left) and
the [Fe VII]6086/\Hb ~(right) emission-line ratios as a function of the
projected
nuclear distance. Same notation as in Figure 1. 
}
\label {fig:fus4}
\end{figure*}

The model results shown in Figs. 2-4 cover most of the data consistently
for the [OIII] 5007+4959/\Hb, [OII] 3727/\Hb, [OI] 6300+6363/\Hb,
[SIII] 9069/\Hb, [SII] 6713+6731/\Hb,
and [FeVII] 6086/\Hb ~line ratios. The first three ratios are related to
the
ionization state of the gas, while the other three are chosen because
they are not well reproduced by pure photoionization models.

The ranges and the spatial distributions of \Fh ~were chosen
phenomenologically
by the fit, particularly, of the  [OIII]/\Hb ~and [SIII]/\Hb ~ratios,
while the ranges
and the distributions of \Vs  ~were chosen  by fitting  the low
ionization
([OII]/\Hb, [SII]/\Hb) and neutral ([OI]/\Hb) line ratios. The data
corresponding
to the high ionization line ratio [FeVII]/\Hb ~are not enough to
constrain the models.

We adopt   a  velocity field decreasing  from the center of the galaxy
outwards.
The observed velocity field is, however, complex, so, two different
shock velocity distributions are shown,
one with a maximum \Vs $\sim$ 700 \kms ~in the central region
(middle diagrams), and another with
a higher  maximum \Vs, up to 1400 \kms ~(bottom diagrams).

Two different lines represent the SD models
in both middle and bottom diagrams.
The thick one corresponds to the calculated models, the thin one to the
models downwards shifted by a factor of 3 (middle diagrams) and of 2
(bottom
diagrams). The shift in the middle diagram is dictated by the
fit of the [OI]/\Hb ~ratios (see Sect 4.1) and in the bottom diagram by
 the [FeVII]/\Hb ~ratios (see Sect. 4.3).
These shifts do not represent lower  abundances
of the elements relative to H, but they indicate that in a multi-cloud
model corresponding to the weighted sum of single-cloud models,
the weights of the SD high velocity  ($>$ 300 \kms) clouds is reduced
(see Sect. 5).
Indeed, by modeling the data on a large scale, this solution may look
arbitrary, particularly
considering faint lines (e.g. [OI]),
and lines affected by the presence of dust (e.g. [SIII], [SII], etc.).
If reducing the weights of the SD models, the fit of all the
emission-lines
is consistently improved, we can conclude that
 this reduction is sound and SD models corresponding
to high velocities in the nuclear region have lower weights.
The  relative weight accounts for
the  relative number of clouds in the conditions determined by the
model, for the dilution factor (i.e., the square
ratio between the distance of the cloud from the galaxy center) and the
distance
of the galaxy from Earth.
Notice that the middle and bottom diagrams in Figs. 2-4
 refer to "high velocities" in the nuclear region of the galaxy
(between 1500-700 \kms and about 300 \kms).
Indeed, Nelson et al (2000, Fig. 5) show that the bulk of cloud
velocities
is  within 300 \kms. So, the reduction of the weights in the SD diagrams
indicates that the number of high velecity SD clouds is small.

Recall that observational data  result from the integration along the
line of sight, which may include different clouds. Thus, a more
realistic fit
should be obtained by a multi-cloud average (see Sect. 5).
A compromise between a consistent picture over a large spatial
distribution
of several emission lines and the precision of the fit must be achieved.

\centerline{\it Observed features}

In several cases there are two or more emission
lines within the passband, so
that different velocities are sampled for the different lines
(Hutchings et al. 1999).
These are the [OII] 3727  and [SII] doublets, as well as [NII] lines
plus
H$\alpha$ at about 6548 \AA.
The [OIII] image covers velocities between -1200 and -860 \kms.
High velocity material in this velocity range is much fainter than the
main bright clouds showing low velocities.
It is seen on both sides of the nucleus and outside
the main biconical emission regions.
The high velocity gas is weak in H$\beta$ and [OII] lines and seems to
be associated with highly ionized material.
Ionization of oxygen is higher along radial locations on both
sides of the nucleus. Generally, there is an association
of high velocity clouds with high ionization gas, however, there are
regions
of high ionization (mainly to the E side) with no known high
velocity gas.
Nelson et al. (2000) claim that the high velocity components
generally account for a small
fraction of the total flux in the [OIII] emission lines.
In several cases they find clouds with multiple velocity
components.

\subsection{Oxygen Lines: [OIII], [OII], and [OI]}

The results for [O III]/\Hb ~and [O II]/\Hb ~are shown in Fig. 2.
Different models are selected to fit the  data as expected from the
multiple structure of the observed lines  (Kaiser et al. 2000). As shown
in
Fig. 2 a, the data   are well reproduced by composite RD models,
with the intensity of the ionizing radiation \Fh ~decreasing
by more than two orders of magnitude from the center
towards the outskirts of the nuclear region.

Focusing on [OIII]/\Hb, RD clouds (Fig. 2 a) with velocities of 200-300
\kms
~contribute preferentially in the inner 1 arcsec region, while those
with
velocities of 100 \kms ~appear  at larger distances from the center, in
agreement with Winge et al. (1999).
The contribution to [OIII]/\Hb ~line ratios from RD clouds with \Vs
$\sim$ 500
\kms ~is small.

Regarding the SD clouds (Fig. 2 b and c), the high velocity ones (\Vs
$>$ 500 \kms) are responsible
for only a few percent of the central region emission, while a larger
contribution
at 3'' from the center comes from the low velocity clouds (\Vs $<$ 500
\kms).

The situation is markedly different regarding [OII]/\Hb ~(Figs. 2 d, e,
and f)
as it appears dominated by SD models. The trend of [OII]/\Hb
~is nicely explained by SD models with
 \Vs ~between $\sim$ 400 and 700 \kms ~(thick line).

Recall that the contribution of high velocity clouds to the [O II]/\Hb
~line ratio is due to the diffuse radiation generated at the high
temperature
post-shock zone  reaching the low ionization zone.
 In order to illustrate this, the distribution
of the temperature as well as  the fractional abundance of oxygen ions
downstream are plotted in Fig. 5 for an SD model with \Vs = 700 \kms.

\begin{figure}
\psfig{figure=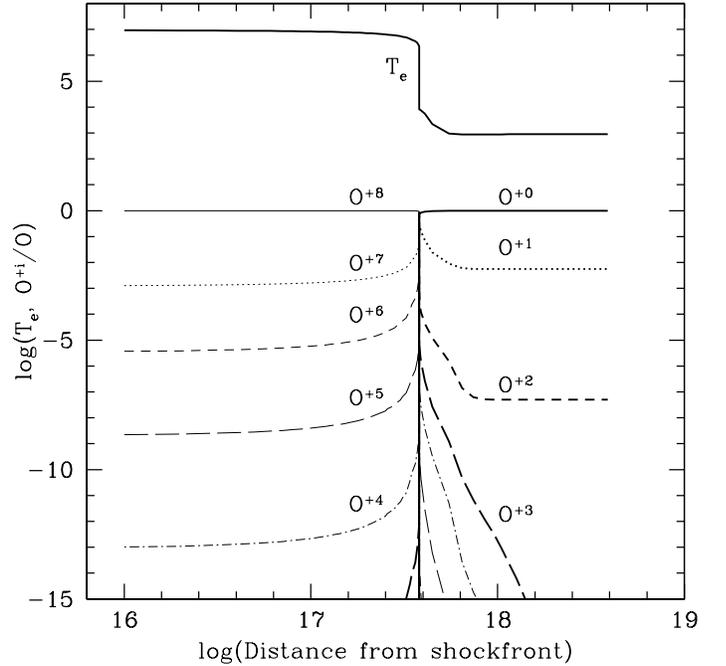,bbllx= 20pt,bblly=
100pt,bburx=750pt,bbury=700pt,width=12.0cm,clip=}
  \caption{The distribution of the electron temperature and of the
ionic fractional abundance of oxygen downstream  for a shock-dominated
model (\Fh=0) with \Vs=700 \kms and \n0=700 \cm3.
The shock front is on the left.
}
\label {fig:fus9}
\end{figure}

Notice, however, that
explaining the [OII]/\Hb ~line ratio only with SD clouds may
not be acceptable in a global picture for the NLR, since RD
clouds must be present and contributing to other emission lines.
Indeed, the reduction by a factor of 3 is not only dictated by the fit
of the [OI]/\Hb
~data (Fig. 3 b), but is consistent with a general scenario.

Regarding the [OI]/\Hb ~line ratio,
notice that SD models  overpredict the observed
[O I]/\Hb ~data
by a factor of $\sim$ 3. This indicates  that SD  models with
 \Vs $\sim$ 700 \kms ~in the central region must be taken with a lower
weight
in an eventual averaged model.
The shift of the SD models dictated by the [OI]/\Hb ~ratios is
consistent with all the other lines.

\subsection{Sulphur Lines: [SIII] and [SII]}

To further investigate NGC 4151 line emission we address a problem
raised  by
Kraemer et al. (2000), namely, the  overprediction of the
[SIII]9069+9532
/\Hb ~line ratio by photoionization models.
This ratio is shown in Fig. 3.

 Different clouds with different
\Vs ~coexist in the center. The larger values can be fitted by
RD models with \Vs $\geq$ 500 \kms.
Notice, however, that RD models with \Vs = 500 \kms ~and log \Fh $\leq$
10.5
overpredict the data in the
region  beyond 1"  at both sides of the center. Therefore, for
consistency, they were taken out from all figures.

There is no contribution from SD clouds to the [SIII]/\Hb ~ratio
in the central region (Fig. 3 e, f).
 On the other hand, a strong contribution
to the [SII]/\Hb ~line ratios  comes from the
SD clouds with rather high velocities (700 \kms) in the central region,
and a not negligeble contribution of SD clouds with \Vs = 1000 \kms
~(Fig. 4 c), as for the case of [OII]/\Hb.
Here too the signature of the diffuse radiation generated at
the high temperature post-shock zone is seen.

From the [SII]6717/6731 diagrams of Nelson et al (2000, Fig. 10), the
gas is rather
clumpy and  also indicates a decrease in density with distance.
The data in the outer region ($>$ 2") are well explained by RD models 
with \Vs = 100 \kms ~and \n0 = 100 \cm3, leading to [SII] 6716/[SII]
6730 $\geq$ 1. 
Density downstream decreases for lower shock velocities and lower
preshock
densities (see Contini \& Aldrovandi 1986), therefore,
[SII] 6716/[SII] 6730 $\geq$ 1 is expected in the regions farther from
the center.
In fact, we have found that velocities generally decrease with distance
from the
center. However, some ratios of about 0.5 are given by Nelson et al. 
They are supported by the fact
that some models with \Vs ~of about 500 \kms also fit the data
beyond 1" (Figs. 4b and 4c).

\subsection{The [FeVII] line}

Fe coronal lines are usually underpredicted by pure photoionization
models. In the case of NGC 4151, [FeVII]/\Hb ~is underpredicted by a
factor of $\sim$ 3 by Kraemer et al. (2000) and by Alexander et al.
(1999).  To
explain the discrepancy, an overabundance of Fe is often suggested,
which is somewhat surprising since a fraction of Fe may
 be locked in grains, although sputtering is strong
for small grains and high velocity shocks.

 [FeVII]/\Hb ~values for P.A. = 221$^o$ are shown in Fig. 4 d, e, and f.
The available
data are scarce, hence further constraints to the models are really not
possible.  The models used to fit the other lines are compared with
the available data. As seen in Fig. 4 f  high velocity models
overpredict the [FeVII]/\Hb ~line ratio.
A better fit is obtained reducing the weight of the high velocity
models by a factor of 2. This reduction does not change our
discussion above concerning the other lines,
because models with \Vs=1400 \kms contribute by
no more than 20 \% to [OII]/\Hb ~and [SII]/\Hb ~line ratios.
 As we will see over the next sections, a low contribution from high
velocity clouds is also needed to explain the IR lines (\S 6)
and the continuum SED (\S 7).
Notice that the [Ne V] 3426 lines, also coming from the
high ionization zone in the clouds, are overestimated by high
velocity models (see \S 5).

\section{The Ne and Si infrared lines}

 The ISO coronal line data reported by  Sturm et al.
(1999) are used to further constrain the  models.
Since the ISO aperture are of the order of several
arcminutes, only  the coronal lines are used.
Lines from lower ionization stages may include the contribution from
star
forming regions in the galaxy. Nevertheless,  they are  included in the
discussion in order to complete the data set with information from
different ionization stages.

Two  Ne coronal lines are observed : [NeVI] 7.6 and [NeV] 14.3 in
addition to
two other strong lines, [NeIII] 15.5, and [NeII] 12.8.
Regarding Si, [Si IX] 3.94, [SiVII] 2.48, and  [SiII] 34.8 are detected.

From all the models plotted in Figs. 2 to 4, we select those
which could better reproduce the IR line ratios. The corresponding input
parameters are listed in Table 1, as well as the weights (W(Ne))
adopted for the best average model (AV).  The weights are relative to
that of model 2.

The weights adopted for the Si lines (W(Si)) are, however,  different
because they
account for Si depletion from the gas phase (and included in grains),
which is different in each model.
To account for the Si coronal line emission, Si/H has to be depleted
by a factor of about 15.
 Moreover, the different depletion of Si
 in the different clouds may be indicating that
the  dust is not homogeneously distributed in the emitting clouds
(see Kraemer et al. 2000).

Notice that the weights are very different for different models because
they must
compensate the difference by many orders of magnitude of the line
intensity fluxes, in order to obtain similar contribution to the
emission
lines from different models (CV01).
SD and RD clouds correspond respectively to
models 2, 3, 4, 6, and  models 1 and 5.

\begin{table*}

\centerline{Table 1}

\centerline{The models for the infrared lines}

\begin{tabular}{l ll llll} \\ \hline\\

\ model & 1 &2&3&4&5 &6 \\
\hline\\

\ \Vs (\kms) & 100& 300&500&500&500&1500\\
\ \n0 (\cm3) & 100& 300&300&300&300&300\\
\ log \Fh    & 11& -&- & -&12.7&-\\
\ D ($10^{17}$ cm)& 100&100&10&100&100&1\\
\ W(Ne)& 5.(-4)& 1.&6.5(-4)&7.9(-4)& 2.(-6)&2.9(-8)\\
\ W(Si)& 0.114& 1.   & 1.4(-3)&1.43(-3)&1.4(-4)&1.6(-6)\\
\hline\\

\end{tabular}

\end{table*}

The observed IR emission-line flux as well as the ratios of the
calculated to observed
lines are listed in Table 2.
This table is organized as follows:
(a) Line fluxes (row 1);
(b) the results corresponding to the best fit model of
Alexander et al. (1999)  (row 2);
(c) single-cloud results (rows 3 to 8);
(d) the weighted average model AV (row 9);
(e) the contribution to the IR emission lines from each single-cloud
model (rows 10 to 15).

Actually, the line and continuum spectra must be modeled consistently,
so, the results
of the AV model in Table 2 are cross-checked by the results of the
continuum SED in Sect. 7,
until the best tuning for both is achieved.

\begin{table*}
\centerline{Table 2}

\centerline{The infrared lines relative to the observation data}

\begin{tabular}{l c cc c cccl} \\ \hline\\

 & [NeVI] & [NeV] & [NeIII] & [NeII] & [SiIX] & [SiVII]& [SiII] &
d$^2$/R$^2$\\
\  fluxes (obs)$^1$ & 7.90 & 5.50 & 20.7 & 11.8 & 0.41 & 1.2 & 15.6&- \\

\  model (calc/obs)$^2$ & 0.5 & 1.8 & 0.7 & 0.5 &0.9 & 0.4 & 3.&- \\
\ model 1 (calc$^3$/obs)& 2.30&10.3 & 2.39 &0.017 &0.064 & 0.137 &
80.4 & 10$^{11}$  \\
\ model 2 (calc/obs)& 1.20 & 0.40   & 0.50  &  0.90 &   17.6   & 3.90 &
6.6& 10$^9$ \\
\ model 3 (calc/obs) & 1.80 &  0.70 &   6.80  &  11.2 &  205. &   11.1 &

36.7&10$^9$ \\
\ model 4 (calc/obs) &  1.80 & 0.70  &  6.70  &  11.0 &  199.  & 10.6 &
27.7 & 10$^9$ \\
\ model 5 (calc/obs)  &1.98&15.0 & 27.5 &0.92 &3.70 &  5.42
&0.11&10$^{13}$ \\
\ model 6 (calc/obs) & 2.05 &1.53& 0.16 & 0.01 & 1.04e5. & 240. &
0.0&10$^9$
 \\
\  model AV (calc$^4$/obs) & 1.30 & 1.24& 1.19& 0.92 & 1.44 & 0.80&
1.0&- \\
\ model 1 \% &8.70 &  42.8  &  10.3 &  0.09   & 3.04 &  11.9 &  57.3 &-
\\
\ model 2 \% &  88.3 &  32.3 &  42.0 &  97.7   &  72.7  &  29.4 & 41.2
&- \\
\ model 3 \% & 0.01 &  0.0    & 0.04 &  0.08 &  1.21 &  0.12 &  0.33&-
\\
\  model 4 \%  &   0.01   & 0.0   &  0.04 &  0.09 &  1.18 &  0.11  &
0.25&- \\
\  model 5  \%  &3.00 &  24.9  &  47.5  & 2.05 &  21.8 &  58.5 &  0.98
&- \\
\ model 6  \% & 0.0    & 0.0   &  0.00 &  0.0    &  0.06  &  0.0   &
0.0 &-  \\
\hline\\
\end{tabular}

 $^1$ in $10^{-13}$ \erg

 $^2$ Alexander et al.

  $^3$ calculated at the nebula

 $^4$ To calculate the averaged model from the models given in CV01 :

      $\Sigma$$_i$ [(F($\lambda$)/F(\Hb))$_{CV01}$  F(\Hb)$_{CV01}$)$_i$
W(model i)]/
      F($\lambda$)$_{obs}$

\end{table*}

Because the theoretical fluxes are calculated at the nebula and
the observed fluxes are measured at Earth, the ratio of the
square distance of the galaxy to Earth (d$^2$)
to the  square distance of the nebula to the galaxy center (R$^2$)
is given in the last column of Table 2.

From Table 2, one sees that model 2 (SD, \Vs = 300 \kms)
strongly contributes to all the lines.
 Gas  is heated to temperatures between 1.3 $10^6$ K and 3.7 $10^6$ K
for SD clouds with  \Vs ~of 300 \kms ~and 500 \kms, respectively.
Radiation dominated model 1 contributes  mainly
to [NeV] and [SiII] and model 5 to [NeIII] and [SiVII].
Model 5 is associated with a very high ionizing flux: 5
$10^{12}$ (cm$^{-2}$ s$^{-1}$ eV$^{-1}$ at the Lyman limit).

The high velocity clouds (model 6, \Vs = 1500 \kms)
do have a low contribution. This  contribution must remain low as to
prevent the increase of  [SiIX] flux far beyond its observational value.

\section{The UV line ratios}

Ultraviolet  lines may provide   important information about their
origin in a gas either heated by  shocks or ionized by a
strong radiation.
The most significant lines are NV 1240, CIV 1550, and HeII 1640.
Other strong UV lines are generally blended (OIV, SiIV 1402,
CIII], SiIII] 1909, etc), and  have not been used in the modeling.
Because the reddening correction in the UV is important, we use
the  NV/CIV and HeII/CIV line ratios instead of intensities
relative to \Hb.

The UV data correspond to  P.A.= 221$^o$ observations.
Model results are presented in
 Fig. 6 where South-West (SW) data
(filled triangles)  are separated from those form the  North-East (NE)
region
(open triangles).

\begin{figure}
\psfig{figure=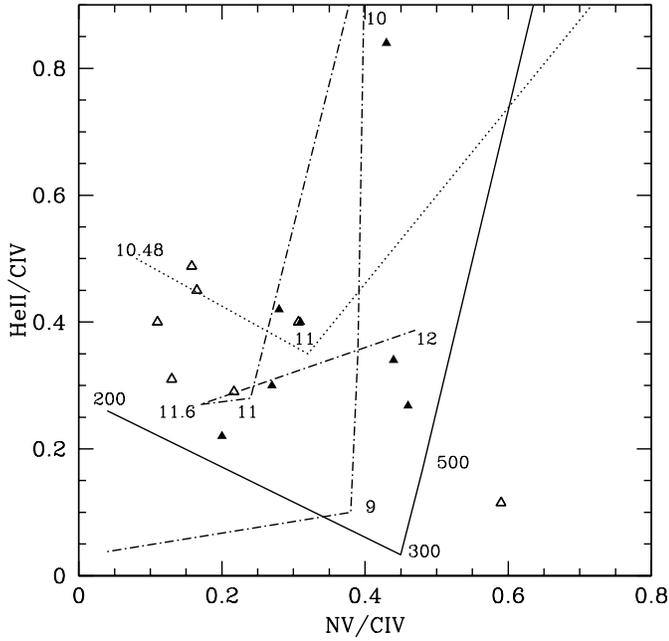,bbllx= 20pt,bblly=
100pt,bburx=750pt,bbury=700pt,width=11.5cm,clip=}
  \caption {He II 1640/C IV 1550 versus N V 1240/C IV 1550. 
SW UV data and NE UV data correspond, repectively,  filled and empty
triangles.
 The RD results  for \Vs = 100 \kms and \n0=100 \cm3 (dotted line),
and \Vs=200 \kms and \n0=200 \cm3 (dash-dotted line) are labelled
by the values of log(\Fh), and the SD results (solid line) by \Vs ~in
\kms.
}
\label {fig:fus5}
\end{figure}

The theoretical results indicate that the NE UV data
can be explained by low velocity RD clouds ($<$ 300 \kms) reached
by an ionizing flux \Fh $< 10^{11}$ units, while the SW UV data
 come from the RD clouds reached by a stronger radiation field,
with some contribution from SD clouds with \Vs $>$ 500 \kms. However,
these
results are only indicative
(see the discussion  by Kraemer et al. 2000).
It is well known that the HeII 1640 line is
strongly dependent on the spectral index of the ionizing radiation,
and the NV/CIV ratio depends on the adopted N/C abundance ratio.
Regarding the HeII lines, if $\alpha_{UV} <1.5$ the theoretical results
may
vertically shift to the upper part of the diagram, and the derived
velocities would be larger.
Regarding the  N/C abundance ratio, the cosmic values is adopted.

For consistency, we present in Table 3 the observed/calculated
UV-optical
 line ratios obtained with the same AV model as for the IR lines
(see Table 2).
To avoid the uncertainty of the dilution factor the ratios of the
AV model  are calculated between the line ratios to \Hb.

\begin{table*}
\centerline{Table 3}

\centerline{The UV-optical lines}

\begin{tabular}{l c cc c cccc} \\ \hline\\

 & [NeIV]2423 & [NeV]3426+& [NeIII]3869+&HeII 4686 &  [OIII]4363 &
[NII]6584+ \\
\  fluxes (obs)$^1$ & 1.77 & 1.56 & 1.81 & 0.31 & 0.35& 2.16 \\
\  model (calc/obs)$^2$ &0.4  & 1.5 & 0.8 & 1.1 & 0.7& 0.8  \\
\  model (calc/obs)$^3$ & -   &  -  & 0.5 & 1.1 & -  & 1.8  \\
\ model AV (calc/obs) & 1.3& 1.5 & 1.5 & 1.1& 1.6&1.4  \\
\hline\\
\end{tabular}

 $^1$ in $10^{-15}$ \erg (Kraemer et al 2000, Table 1, 0".1-0".3)

 $^2$ Kraemer et al. (2000, Table 1, 0".1-0".3)

 $^3$ Alexander et al. (1999)

\end{table*}

The fit is within 1.6 and the trend is always to overestimate the
values.
This is consistent with reduction of the SD models which was adopted in
Figs. 2-4  b, c, and e, f .

\section {The continuum}

To further constrain our model for NGC 4151, the continuum SED
is analysed and modeled using  the
composite model derived from the Ne and Si lines in \S 5.

\centerline{\it Observations}

 Gamma-rays to the optical continuum emission
 is taken from
 the  1993 December  multiwavelength
monitoring campaign of NGC 4151 (Edelson et al. 1996).
The source  was near its  peak historical
brightness  during this campaign and showed the strongest variations of
its
continuum emission  at
medium energy  X-rays ($\sim$1.5 keV) with amplitude variations of
$24\%$, weaker variations (6$\%$)  at the gamma-ray energies, decreasing
variation from ultraviolet (9$\%$) to optical (1$\%$) and finally, not
significant variation at the soft X-rays (0.1-1 keV).

Assuming  that the continuum emission variations are within a few
percent level beyond optical wavelengths, the 1993   data are
combined with
IR  data  taken at different epochs.  Since the optical aperture used by
Edelson et al. (1996) is 12x3
arcsec,  near -IR fluxes  from  an aperture size of  10  arcsec
is  adopted when possible. Sources are as follows:

The data for the near-IR come from the NASA Extragalactic Data
(NED), J, H and K data are taken from Balzano \&
Weedman (1981) and L and M data  from McAlary et al (1979).
The 10 microns emission is from Lebofsky \& Rieke (1979) and
corresponds to an aperture size of 6 arcsecs.
In addition, to trace  the non-stellar contribution at the
near-IR,  Kotilanien's et al. (1992) data  within a 3 arcsec aperture
are  also considered for comparative purposes.

For the far-IR region  between 16 and 200 microns,  ISOPHOT data from
Perez-Garcia et al. (1998) are taken. Due to the large aperture
size used in ISOPHOT, those data are  integrated over the complete
galaxy.

For the radio data, values at 1.4 GHz,
4.85 GHz, and  408 MHz  were at the NED, and come  from Becker
et al. (1995),
Becker et al. (1991), and
 Ficarra, Grueff
\& Tommasetti (1985), respectively, while data at 8.4 and 5 GHz
are taken from Pedlar et al. (1993).

Compared with other Seyfert galaxies, NGC 4151 is relatively weak in
X-rays
with L(2-10 keV) $\sim 7 \, 10^{42}$ erg s$^{-1}$ (Weaver et al. 1994).
The  soft X-ray emission is extended (Morse et al. 1995).
Recent Chandra data resolve
up to $\sim $ 70 $\%$ of the 0.4-2.5 keV emission (Ogle et
al. 2000). This emission  appears to be associated with
the optical narrow line gas (NLR) extending asymmetrically
to the South-West of the nucleus.

\bigskip

\bigskip
\centerline{\it On the nature of  the soft X-ray emission}
\bigskip

Both, Weaver et al. (1994) and Ogle
et al.(2000) provide similar  estimates for the
plasma pressure of the hot
gas of  about $ 3 \times 10^7 cm^{-3} K$. This is one
order of magnitude larger than the pressure derived from the cold NLR
gas (Penston et al. 1990). The difference  strongly argues against  the
soft
X-ray emission being associated with the NLR confining medium (Weaver
et al. 1994,  Ogle et al. 2000).

Weaver et al. (1994) also found that the Fe K edge
energy in their 0.4 - 11 keV  data  (from the BBXRT mission)   is
inconsistent with an origin  in a  gas
with the same ionization parameter as the low-energy absorber.
This allows the authors to rule out a line-of-sight ionized absorber
as the sole source of the soft X-ray excess in NGC 4151
(see also  Contini \& Viegas 1999 for the case of NGC 4051).

George et al. (1998) propose a model for the 0.2-10 keV
spectrum where the   underlying power law nuclear component is
partially  absorbed by an ionized absorber and partially
scattered. However, it requires  an additional
component due to Bremsstrahlung thermal
emission from an extended photoionized gas at  T$\sim$ 6 $10^6$K.

Recent Chandra data provide direct evidence of X-ray line emission gas
at T $\sim 10^7$ K. The strength and ionization potential of the
X-ray narrow emission lines indicate a composite spectrum in which
both photoionization and shocks are at work (Ogle et al. 2000).
The fair spatial  association between
the central NLR optical region and the soft X-ray emission indicated
that both mechanisms are contributing to the multi-wavelength SED
and to the line spectrum of NGC 4151 (cfr. Komossa  2001).
Taking into account these observational constraints, the
origin of this soft X-ray excess is evaluated in the next section
assuming  high velocity models dominated by shocks.

\centerline{\it Model Results}

NGC 4151  shows the characteristic of Seyfert galaxies rather than
 starbursts. Therefore, we consider that the continuum
SED is Bremsstrahlung radiation from clouds ionized and heated
by the radiation from the AC and by shocks.
We   refer to previous studies (e.g Contini \& Viegas-Aldrovandi 1990,
Contini \& Viegas 1991, Viegas \& Contini 1994, Contini \& Viegas 2000,
etc)
on this subject which could lead to a better understanding of the
results.
Particularly, dust and gas are coupled entering the shock front and
mutually heat each other. The grains are  collisionally heated to
the highest temperatures ($\geq$ 300 K) leading to emission in the
near-IR.
On the other hand, heating by the central radiation is not efficient
enough, so dust in the radiation dominated zone does not reach such high
temperatures.

Before discussing model results, let us notice that below the Lyman
limit, up to say 0.2 keV, we have very little observational information:
this is the "unknown window" where observations are difficult because
of the heavy absorption by our Galaxy.  However, in the 0.2 to 2 keV
region, NGC 4151 shows extended  emission, spatially
correlated with the optical NLR gas.  If due to Bremsstrahlung,
the  tail of this X-ray emission should somehow show up in the far UV.
However, no  detected extended emission appears in the UV HST images
of NGC 4151 (Boksenberg et al. 1995). Therefore,  the emission from the
NLR gas should be lower than that of the UV nucleus of NGC 4151.  We
 will use this observational fact as a constraint in the proposed
modeling in the sense that the observed
 nuclear UV emission  can   only be
associated with model emission within a nuclear region of outmost ~3 pc
size.

Observational data and model results are plotted in Figs. 7 and 8,
which refer to models and relative weights adopted to fit the Ne
infrared
lines (Table 1, row 5).  Notice that Bremsstrahlung peaks at high
frequencies depending on \Vs (see Viegas \& Contini 1994).

The large geometrical thickness of the emitting clouds and the high
densities downstream due to compression (n/\n0 $\geq$ 10, depending on
\Vs, \n0, and \B0) lead to high optical thickness of the emitting gas
with column densities of the order of 10$^{21}$-10$^{24}$ cm$^{-2}$.
This is within the range of  column densities estimated from
the X-ray data, $\sim$ 5 10$^{22}$ cm$^{-2}$ (Weaver
et al. 1994, George et al. 1998,  and Ogle et al. 2000).
Accordingly,  the emitted radiation gets absorbed in the cloud itself
between $\sim$ 13.6 eV and 500 eV, hence the emission gap in this
region  shows up  in Figs. 7 and 8.  The high velocity model 6
has however a lower column density ($<$ 10$^{21}$ cm$^{-2}$)
because is radiation-bounded.
In fact, due to strong compression, the temperature rapidly
decreases downstream at a distance $< 10^{16}$ cm from the shock
front.  Thus,  the contribution of model 6 to the SED extends
into the far UV range (Fig. 7 long-dashed line). However, the
 X-ray data in the frequency range  $ 10^{15}$ and $10^{17}$ Hz impose
a
limit on the relative constribution of this model 6 (see below).
All together,  the  fit of the continuum at the UV range is
obtained with model 2.
The dust-to-gas ratio that  characterizes this model is
d/g = $10^{-14}$. The other models (models 3, 4, 5, and 6) underpredict
the data.

\begin{figure}
\psfig{figure=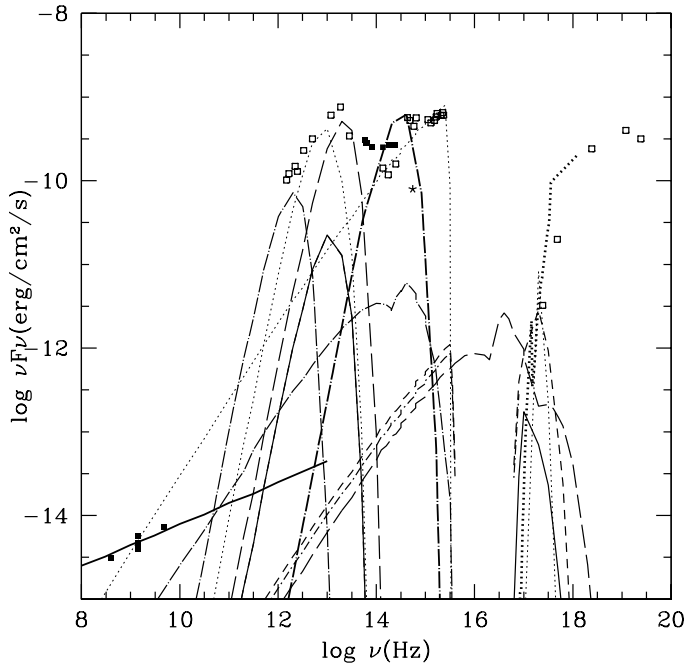,bbllx= 22pt,bblly=
100pt,bburx=700pt,bbury=700pt,width=11.0cm,clip=}
  \caption {Different components contributing to the SED of the
continuum of NGC 4151,
corresponding to the results of several models, scaled following the
weights used to fit
the IR lines. The models are the following:
1 (dash-dotted line), 2 (dotted
line),
3 (short dash-dotted line),4 (short-dashed line), 5 (solid line), 6
(long dashed line).
The flux from the AC in the far X-rays, which is seen through the
clouds,
is represented by a thick dotted line, while the continuum from the old
stellar
population is indicated by a thick long dash-dotted line.
The theoretical results are compared to observational data taken from
the NED
(filled squares)  and  from other sources described in Sect. 7 (open
squares).
}
\label {fig:fus7}
\end{figure}

\begin{figure}
\psfig{figure=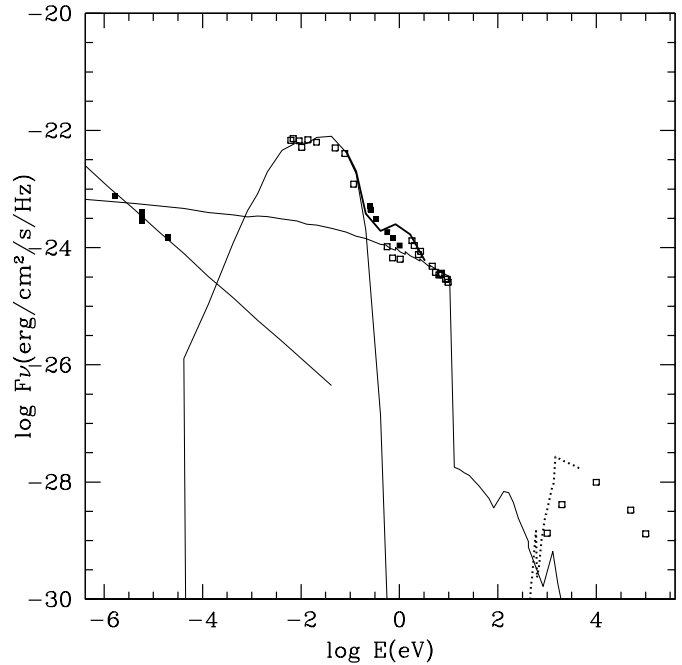,bbllx= 22pt,bblly=
100pt,bburx=700pt,bbury=700pt,width=11.0cm,clip=}
  \caption {
 The continuum spectrum of NGC 4151. The thin solid lines
correspond to the
summed theoretical contributions from dust emission, gas Bremsstrahlung
and radio emission as shown in Fig. 7. The thick solid line
corresponds to
the summed contributions from dust emission,  gas Bremsstrahlung and the
old
stellar population continuum, while the dotted line corresponds to the AC
continuum seen
through the clouds. The observational data  are taken from the NED
(filled squares)  and  from other sources described in Sect. 7 (open
squares).
}
\label {fig:fus8}
\end{figure}

The summed SED which better explains the observations is shown in Fig.
8.
The main components are summed separately: radio emission, reradiation
by dust in the IR, gas Bremsstrahlung, and the flux from the active
source.  Moreover, as was found for many Seyfert galaxies (see Contini
\& Viegas 2000)
 a black-body with temperature of $\sim$
3000 K is used to roughly represent the optical continuum due to the
stellar population (thick solid line).

 Notice that we are interested in the bulk contribution of the stars
rather than
the detailed shape of their continuum emission. Thus, although we could
use 
the  spectrum of an early-type galaxy or the spectrum of the central
region of 
a Sa galaxy to get a more realistic fit of the stellar contribution, 
it would not change our conclusions.

For instance, Malkan \& Filippenko (1983) find that about 1/10 of the
emission at 5500 \AA
and $\sim$ 1/3 at 9000 \AA are due to stars (in aperture of 10",
just the same aperture of the optical and IR data used in this paper),
namely they estimated the stellar emission to be 14 mJy at 5500 \AA.
We have added the corresponding value to
 Fig. 7 (star). Accounting for the errors in the observations
and models, notice the good agreement 
(within 30\%) with the black body emission curve.  

The bump in the IR  is rather wide, representing the
sum of  contributions from  different models. In previous papers
(e.g. Contini et al. 1999b, Appendix A) it was explained that the bump
in the IR due to each model depends on the shock velocity.
Particularly, the high velocity model (model 6) has its maximum at the
near-IR.  To fit the near-IR data,  a high d/g  (2 $10^{-12}$) is
assumed
in model 6,
indicating that high velocity material is rather  dusty.
A shock velocity of  1500 \kms  ~produces a post shock region at a
temperature of 3.4 $10^7$ K, in agreement with the  hot plasma
temperatures derived from the X-ray data.
indicating that high velocity material is rather  dusty.

Shocked clouds with \Vs $\sim$ 700 \kms ~produce lower  temperatures,
$\sim$ 10$^6$ K, which are still within the range of values estimated
from
 X-ray data (e.g.  George et al. 1998).
These velocities are also revealed by  optical  lines (\S 4).

Yet, the contribution of a model with  \Vs = 1500 \kms ~in the SED
should be taken with a relative small weight ($\leq$ 2.9 $10^{-8}$),
because the
data in the soft X-ray range constrain the model. As a consequence,
the contribution of this model to the IR lines is very low. Thus,
a signature of these high velocity clouds is not expected to
be seen in the infrared emission-line spectrum.

In the radio range,  Pedlar et al. (1993) derive an  average spectral
index
for the entire source of $\alpha$ = -0.87 from their 8.4 and 5 GHz
data. They interpreted the emission as due to thermal free-free
emission from the NLR.
In the present modeling, the radio data  are instead
well explained by synchrotron radiation
with spectral index  -0.75, generated by
 Fermi mechanism at the shock front (Bell 1978).
Notice that self-absorption of the Bremsstrahlung
emission at these radio wavelengths may be important, hence its
contribution
is  negligeble (see, for instance Contini et al 1998b).

\section{Discussion and Conclusions}

In this paper we have modeled the narrow emission-line
and  continuum spectra of NGC 4151 with particular attention to the
large range of velocities indicated by the line profiles.

The analysis of the line profiles is complex and
clouds in many different conditions can contribute to each line.
By modeling the NLR to analyse the high spatial resolution data
(Nelson et al. 2000, Kraemer et al. 2000), it is found that the
contribution
of each cloud to a given line may show a large variation from line to
line. 

In the central region high velocity clouds  revealed by the observations
(Hutchings et al. 1999, Winge et al. 1997, etc) are shock dominated
(SD).
There is a strong contribution of SD clouds with shock velocity \Vs
$\sim$ 700
  \kms ~to [OII] 3727/\Hb, and [OI] 6300/\Hb.
Radiation dominated (RD) clouds  with \Vs $\sim$ 500 \kms ~are necessary
to explain  [SIII]/\Hb ~and [OI]/\Hb, while clouds
  with low \Vs ~(100-200 \kms)  and reached by a relatively
strong ionizing radiation (log \Fh $\sim$ 11)
contribute to the [OIII]/\Hb, [OII]/\Hb,  and [SII]/\Hb ~line ratios.

Beyond 2" from the center, RD clouds, photoionized by a weaker
radiation flux are responsible for the observed line emission.

\bigskip

Quantitative modeling of the  Ne and Si infrared lines
from different ionization levels  leads to  more  results.
Notice, however, that the  observations give the integrated flux
from all the galaxy, so that it is not possible to understand
which type of clouds prevail in different regions.

SD clouds with \Vs = 300 \kms strongly contribute to  all Ne lines
corresponding
to different ionization levels (Table 2), while
RD clouds with velocities of 500 \kms ~and reached by a strong radiation
flux (log \Fh = 12.7)
contribute particularly to the  [NeIII] and [SiVII] IR lines.
Due to the high postshock temperature, high velocity clouds
(\Vs = 1500 \kms) would overpredict the [SiIX] 3.94 line.
Therefore, the contribution of these clouds is low.
A large contribution  to [NeV] and [SiII] lines
come from low velocity (100 \kms) clouds reached by an
ionizing radiation characterized by log \Fh = 11.

Modeling of the IR lines also shows that
silicon   is depleted by a  factor of 15
because  included in dust grains, while N/C abundance ratio is found
compatible with cosmic values from  the modeling of the UV lines.

Comparing the results of the present work with those of other authors,
notice that our results are slightly better.  Indeed, better fits
could be obtained with a different choice of weights. However, the
results presented in Tables 2 and 3 are consistent with the fit of the
continuum SED (\S 7) and were chosen accordingly.  More particularly, it
seems that the three Ne emission lines are higher than observed by
factors of 1.3 - 1.5. This may be an indication that the Ne abundance
we have used is a factor of 1.3-1.5 too high.
So, decreasing the Ne/H abundance by
1.3, we get a better agreement.  The HeII 4686 line is slightly higher,
but
this is mainly due to the power law above 54.4 eV, which we (and the
others) could have taken somehow too flat.  The only problem is the
[OIII]4363 line. Because oxygen is a coolant, changing the O abundance
may
not solve the problem.  However, a factor of 1.6 higher is still
within the limit of a factor of 2 proposed by Alexander et al., who do
not give their results for [OIII] 4363.

Notice that we based our models on IR lines and we compared with
Alexander et al. (1999).  So, we should refer only to them.  The results
of
Kraemer et al in row 2 of Table 3 are better than ours only for [OIII]
4363 and [NeIII].  Actually, they come from observations in the region
between 0.1" and 0.3" SW, whereas we average on all the regions.

\bigskip

The analysis of the continuum SED  leads to the following results:

The high velocity material is very dusty
(d/g$\sim 10^{-12}$).
The radio emission is synchrotron created by Fermi mechanism
at the shock front.
Moreover, shocks are important to explain the soft X-rays:
shocks with velocities of  1500 \kms  ~produce a post shock region at a
temperature of 3.4 $10^7$ K.
This value is  in agreement with  the temperature found by Weaver et al.
(1994) and Ogle et al. (2000) for  a non-equilibrium plasma from
X-ray observations, while
 clouds with \Vs $\sim$ 750 \kms ~produce temperatures  of
8. $10^6$ K (George et al. 1998).

\bigskip

An estimate of several physical quantities applying to the central
AGN region can be derived from the proposed modeling.

If d is the distance  from Earth  (19.8 Mpc) and d$^2$/R$^2$ = 10$^9$
(see Table 2), R $\sim$ 0.626 kpc  is the average distance of
the emitting nebula to the AC.
Adopting an average downstream density n=10$^5$ \cm3, and D=$10^{18} -
10^{19}$ cm,
the calculated mass is  M $\sim$ 3.5 $10^{8-9}$ ff \msol,
where ff is the filling factor which is likely to be less than unity.
In this case our estimation of the NLR mass is lower than the values
quoted
by Ulrich (2000) for an emitting region closer to the center, i.e.
10$^9$ \msol within 40 pc and 5 $10^7$ \msol
within 12 pc (0.15").
The corresponding calculated central source luminosity is about 1.6
$10^{43}$
erg $\rm s^{-1}$.
The average kinetic energy of the NLR is about 3 $10^{56}$ ff ergs,
assuming an average velocity of the NLR clouds of 300 \kms.
This would imply that the time scale for the AGN phase in NGC 4151
should be
larger than 5 $10^5$ years.

Summarizing, we provide a self-consistent modeling of the
multiwavelength line spectrum and SED of the nuclear region of NGC 4151.

The modeling is based on the coupled effect of shocks and
photoionization operating in the narrow emission line gas. As such, it
implies the existence of emitting clouds with
velocities and densities in a large range.  An emitting-cloud
distribution in the
velocity/density space is precisely what is revealed by the HST
observations of this galaxy.  This together with the  fair
fitting of the line and continuum spectra obtained in this work
reinforces
 our hypothesis that shocks and photoionization are effectively coupled
in
the NLR of AGN.

\begin{acknowledgements}
We are grateful to the referee for many helpful comments.
This paper is partially supported by the Brazilian agencies:
CNPq (304077/77-1),
PRONEX/Finep (41.96.0908.00), and FAPESP (00/06695-0).

\end{acknowledgements}

\end{document}